\input{aipcheck}

\documentclass[
    ,final            
  ]
  {aipproc}
  
\layoutstyle{8x11single}

\newcommand{\apj}{ApJ}

\newcommand{\apjs}{ApJS}

\newcommand{\mnras}{MNRAS}
\newcommand{\aj}{AJ}

\begin{document}

\title{Astrophysical Interplay in Dark Matter Searches}

\classification{95.35.+d; 98.10.+z} 

\keywords      {cosmology: dark matter; Galaxy: kinematics and dynamics; elementary particles}

\author{Louis E. Strigari}{
  address={Kavli Institute for Particle Astrophysics and Cosmology, Stanford University, Stanford, CA 94305}
}

\begin{abstract}
\par I discuss recent progress in dark matter searches, focusing in particular on how rigorous modeling the dark matter distribution in the Galaxy and in its satellite galaxies improves our interpretation of the limits on the annihilation and elastic scattering cross sections. Looking forward to indirect and direct searches that will operate during the next decade,  I review methods for extracting the properties of the dark matter in these experiments in the presence of unknown Galactic model parameters. 
\end{abstract}

\maketitle

\section{Introduction}
\par It is now the beginning of the decade in which we expect incredible progress in the field of particle dark matter detection. Through a variety of indirect searches, direct searches, and collider searches around the world there has already been a series of exciting developments in particle dark matter searches over the course of the past several years. As progress in this field will continue to accelerate over the next several years, the impetus to understand how to best interpret the results in the larger context of cosmology and high energy physics will become stronger. 

\par It has now been over four years since the highly-successful launch of the Fermi Large Area Space Telescope (LAT), which is sensitive to photons in the energy band of $100$ MeV to $300$ GeV~\citep{Atwood:2009ez}. The Fermi-LAT has expanded upon our window into the high energy universe, providing a wealth of information about pulsars, active galactic nuclei, and diffuse gamma-ray radiation on the scale of the Milky Way and beyond. In the area of particle dark matter, as I will discuss in more detail below, the Fermi-LAT has proven to be the first experiment to gain sensitivity to the thermal relic scale of the dark matter annihilation cross section of $\langle \sigma v \rangle \simeq 3 \times 10^{-26}$ cm$^3$ s$^{-1}$. Because of the LAT, for the first time we are now able to explore the regime of cosmologically-motivated weakly interacting massive particle (WIMP) dark matter. 

\par Complementing the indirect results from the Fermi-LAT, direct dark matter searches, such as XENON100~\citep{2012arXiv1207.5988X} and the Cryogenic Dark Matter Search (CDMS)~\citep{Ahmed:2009zw}, are continually improving the sensitivity to the WIMP-nucleon interaction rate. These experiments are now beginning to reach into the well-motivated theoretical regime of Higgs-mediated dark matter interactions, and are now beginning to constrain well-motivated regimes of the supersymmetric model parameter space. As these searches, and many others, continue to improve over the next several years, it is certainly fair to say that either WIMPs will be detected, or there will be a paradigm shift in our strategy regarding searches for particle dark matter. 

\par In this contribution I will discuss how our modern understanding of dark matter in the Galaxy, satellite galaxies, and dark matter subhalos impact our interpretation of the results from both indirect and direct dark matter searches. For the indirect searches, I focus primarily on the Fermi-LAT results, highlighting what we are now able to robustly extract about the nature of WIMP dark matter from studies of a variety of astrophysical environments. During the discussion of direct searches, I review our observational understanding of the local dark matter distribution, how it affects modern results, and methods for extracting the properties of the WIMP in the future, even though our knowledge of the dark matter distribution in the Galaxy is still not as precise as we ultimately desire. 
 
\section{Indirect Searches}
\par Examination of the all-sky gamma-ray map in the Fermi-LAT band reveals that the largest source of gamma-rays results from diffuse emission in the Galactic plane due to cosmic ray interactions with the interstellar medium. Gamma-rays are produced via neutral pion decay, bremsstrahlung, and inverse Compton emission~\citep{Abdo:2009mr}. In addition to this diffuse emission, there are 1873 known point sources in the two-year source catalog~\citep{2012ApJS..199...31N}. The majority of the identified point sources are active galactic nuclei, while a smaller fraction of them are supernova remnants, globular clusters, and star-forming galaxies such as M31 and the Magellanic Clouds. Of the detected point sources, less than one percent of them are firmly identified in other wavebands, while approximately 60\% of them are reliably associated with sources at other wavelengths. Because the angular resolution of the LAT is approximately $0.1-1$ degree, depending on energy, the association with sources at other wavelengths is done via spectral or timing information. Interestingly, over 30\% of the Fermi-LAT sources are unidentified in other wavelengths. 

\par Extraction of a potential dark matter signal from the Fermi-LAT data requires an understanding of both point source emission and diffuse emission over more extended regions. To date, searches for dark matter have been undertaken from a variety of sources within the Galaxy and beyond, including dwarf spheroidal (satellite) galaxies, dark matter subhalos that are not associated with stars, and clusters of galaxies. There have also been analyses of diffuse emission from Galactic and extragalactic sources, and searches for gamma-ray lines from the region of the Galactic center and in the diffuse halo. Each of these analyses provides a unique potential signature of dark matter, as well as unique sets of systematics that must be understood. 

\par At this stage, the dwarf spheroidal (dSph) results are most robust dark matter results that have been obtained by the Fermi-LAT. The dSph analysis relies on a unique confluence of results from the fields of Galactic optical astronomy and gamma-ray astronomy, and the searches are now sophisticated to the point where they may be viewed as particle dark matter experiments in the sky. For these reasons, I will review the methodology for obtaining the dSph results, as well as how these results are expected to improve in the future with more Fermi-LAT data, and data from ground based Air Cherenkov Telescopes (ACTs). I will then compare these results to searches for dark matter annihilation from other sources with the LAT. 

\subsection{Satellite galaxies} 
\par There are now nearly two dozen galaxies that are classified as satellites of the Milky Way (see Ref.~\citep{McConnachie:2012vd} for a recent review of their properties). More than have of these have been identified in the Sloan Digital Sky Survey (SDSS), so the majority of these objects are concentrated towards the North Galactic Gap. The luminosities of these systems vary between a few hundred solar luminosities to tens of million solar luminosities. Combining the stellar kinematics of these galaxies with their measured luminosity, they have dark-to-luminous mass ratios of anywhere from tens to even thousands to one~\citep{Strigari:2007at,Strigari:2008ib}. The dSphs that are utilized in the Fermi-LAT analysis contain no detectable interstellar gas, so there is no expected gamma-ray emission from conventional astrophysical sources. Any observed intrinsic emission in these systems would be due to dark matter. 

\par Detailed analysis of the kinematics of these systems over the past several years has revealed that the mass within their approximate half-light radii are well-constrained by the observed data~\citep{Strigari:2007vn,Walker:2009zp,Wolf:2009tu}. The half-light radii for these systems vary anywhere between a few tens of parsecs for the faintest of them, to a few hundred parsecs for the brightest. The angular scale subtended by their respective half-light radii are then in the range $0.1-1$ degrees. This is an important scale, because it implies that both the integral over their density and their density-squared distributions are well constrained on this angular scale. Assuming that the central density cusp is less steep than approximately $r^{-1.5}$, the integral over the density-squared is insensitive to the presence of a core or a cusp in the central density of dark matter. Typical uncertainties on the (log of) the integral of the density-squared, which is the relevant quantity for dark matter annihilation, are approximately 10\%, depending in detail on the numbers of kinematic tracers used to derive the dark matter distribution. The uncertainties are found to be log-normal to a good approximation. 

\par With the integral of the density-squared of the dark matter distributions, $\int \rho^2 dV$, determined from the stellar kinematics, in combination with a model for the WIMP, its annihilation cross section, and its corresponding spectrum of gamma-ray radiation, the flux of gamma-ray can be precisely predicted from the dSphs with high quality kinematic data~\citep{Strigari:2006rd,Strigari:2007at,Martinez:2009jh,Charbonnier:2011ft}. The dSphs with the largest flux are found to be Ursa Major II (32 kpc), Segue 1 (25 kpc), Draco (80 kpc), and Ursa Minor (66 kpc). The former two were identified in the SDSS, and have fractional uncertainties in $\int \rho^2 dV$ still approximately 50\% because they have relatively small samples of stars from which kinematic information can be extracted~\citep{Simon:2007dq,Simon:2010ek}. On the other hand, Draco and Ursa Minor have hundred of stars associated with them, so their fractional uncertainties are of the order 10\%. 

\par In the two-year Fermi-LAT data, examination of the dSphs reveals no excess above the known backgrounds from diffuse Galactic emission, extragalactic emission, and nearby point sources. A joint likelihood with 10 dSphs finds that at the 95\% c.l., WIMPs with mass in the range 10-25 GeV that annihilate to $b \bar b$ and $\tau \bar \tau$ are ruled out~\citep{Ackermann:2011wa,GeringerSameth:2011iw}. This result is additionally robust to the presence of dark matter substructure, which is expected to be not important for the dSphs~\citep{Springel:2008cc,Martinez:2009jh}. This result implies that thermal relic WIMPs that annihilate predominantly through s-wave (velocity independent) interactions in this mass range are ruled out. The significance of this result cannot be overstated, because it is the first time that thermal relic dark matter is robustly being probed via astrophysical observations. In addition to this lack of continuum emission no lines have yet been found in any of these sources~\cite{GeringerSameth:2012sr}. 

\par The present dSph limits lose sensitivity at WIMP masses of approximately 1 TeV. Above this mass scale, ACTs, which have energy thresholds of approximately $100$ GeV and better angular resolution than the Fermi-LAT, are able to complement Fermi-LAT searches~\citep{Essig:2009jx,Essig:2010em}. There are now several ACTs that have studied nearby dSphs~\citep{Aleksic:2011jx,Aliu:2012ga}. Because the exposure of ACTs on the dSphs is significantly less than the Fermi-LAT exposure, the limits on the annihilation cross section are weaker, typically about two to three orders of magnitude above the thermal relic WIMP scale. 

\subsection{Dark subhalos}
\par Cold dark matter theory predicts that, in addition to the dark matter subhalos that host the dSphs,  there may be many orders of magnitude more that do not have any visible stars associated with them (e.g.~\citep{Strigari:2010un,BoylanKolchin:2011de}). These subhalos may in some cases be bright enough to be identified by the Fermi-LAT through dark matter annihilation. Extracting this signal represents a signifiant challenge, because our knowledge of the subhalo distribution in the galaxy (if they even exist) is incomplete and depends on how the cold dark matter mass function is extrapolated to lower mass scales. 

\par In order to get an idea of whether it is now possible to detect dark subhalos with Fermi-LAT, as a first pass, one can ask which of the unidentified point sources mentioned above have a gamma-ray spectrum that is consistent with WIMP annihilation~\citep{Buckley:2010vg}. Even if these sources are consistent with a WIMP spectrum, there are two issues that make a clean interpretation of a WIMP signal difficult. First, there  is significant contamination from astrophysical sources whose spectra mimic the high-energy exponential cut-off that is characteristic of a WIMP spectrum. Second, it is highly unlikely that hundreds of sources would be visible in the present LAT data-- N-body and gamma-ray simulations predict that, for a thermal relic cross section, the number of visible satellites should be not much greater than a few at the most, accounting for detection efficiency cuts~\citep{Ackermann:2012nb}. 

\par Though in principle emission may be detected from an extended dark matter subhalo that is both nearby (i..e within a few kpc from the Solar neighborhood), it is more probable that a dark subhalo that is detected will have a mass greater than approximately $10^7$ M$_\odot$. This is because the mass function of subhalos is $dN/dM \propto M^{-1.9}$, implying that the majority of the total mass in subhalos is locked up in the most massive objects. Accounting for detection efficiency and for the fact that some of the sources may be extended, a search in the one year Fermi-LAT data has revealed that there are no conclusively-detected dark matter subhalos~\citep{Ackermann:2012nb}. Initial promising candidates were eventually correlated with astrophysical sources at other wavelengths. For a $100$ GeV mass WIMP, this null detection corresponds to a limit on the annihilation cross section approximately two orders of magnitude greater than the thermal relic scale. 

\subsection{Galactic center} 
\par In contrast to the dSph analysis, it is not yet possible to make reliable predictions for the flux of gamma-rays from the Galactic center, because there are no direct empirical constraints on the dark matter distribution in this region. Kinematic data are consistent with both cored and cusped dark matter profiles. Further, the diffuse emission from neutral pion decay, bremsstrahlung, and inverse compton is not known to the precision that is required to accurately subtract out the gamma-ray emission that traces back to cosmic rays. In spite of this lack of understanding of cosmic ray induced emission, recently some studies have hinted that large scale features in the diffuse gamma-ray emission towards the Galactic center are consistent with what is nominally expected from a WIMP induced gamma-ray spectrum~\citep{Hooper:2010mq,Abazajian:2012pn}. Further, there have been more recent suggestions of a line feature near the Galactic center~\citep{Weniger:2012tx}. More will be learned from Fermi-LAT data, as well as from HESS-II data, about these potential features in the coming years.

\subsection{Diffuse searches} 
\par If the dark matter has the thermal relic cross section and has a significant s-wave component, it may also be identified over larger angular scales in the Galactic halo. However, extraction of this signal is difficult at present, because it relies on marginalizing over a large number of parameters that describe the diffuse emission from cosmic rays~\citep{Ackermann:2012rg}. Further, there is uncertainty in the scale of the annihilation cross section that these results are testing due to the uncertainty in both the local dark matter distribution and the total mass of the Milky Way. The case is similar for indirect dark matter searches that rely on extracting a dark matter signal from the isotropic background~\citep{Abdo:2010dk}, which has an observed power law with a spectral index of $2.4$~\citep{Abdo:2010nz}. The predictions for the extragalactic radiation component from dark matter annihilation is particularly difficult due to the unknown contribution to the flux from dark matter substructure in galaxies. Complementing these results from gamma-ray searches, it is also interesting to note that searches for diffuse neutrinos are now able to place limits on the annihilation cross section into neutrinos, but the sensitivity is a few orders of magnitude worse than the gamma-ray searches~\citep{Abbasi:2011eq}. 

\subsection{Galaxy clusters} 
\par Several nearby galaxy clusters are promising targets for indirect dark matter searches. For the nearest clusters, because of the appropriate mass and distance factors the emission is predicted to be similar to that from satellite galaxies. Also, several nearby clusters have well-deterimned mass profiles, so it is possible to make detailed predictions for the expected gamma-ray flux in a manner similar to the case of the dSphs. However, unlike the dSphs, there is expected to be significant gamma-ray emission from clusters on the scales probed by the cluster gas distribution~\citep{Pinzke:2010st}; understanding this signal in more detail is required to extract a WIMP signal. 

\par There have been no conclusive detections of gamma-ray emission from clusters with the Fermi-LAT to date~\citep{Ackermann:2010rg}. For a smooth dark matter mass distribution, at $10$ GeV this non-detection constrains the annihilation cross section $\langle \sigma v \rangle$ to less than approximately $10^{-25}$ cm$^3$ $s^{-1}$ for annihilation into $b \bar b$~\citep{Ando:2012vu,Han:2012uw}. However, a more precise determination is difficult due to the uncertainty in the component of the annihilation that results from halo substructure. Indeed for extrapolation down to Earth mass scales, the emission from substructure may increase the smooth flux by about three orders of magnitude~\cite{Gao:2011rf}. The vast majority of this emission is expected from the outer regions of the cluster where the dominant component of the substructure is distributed. Understanding the nature of this extended emission, and separating it out from the less extended emission that traces the gas distribution, will be the most important aspect of gamma-ray cluster analyses going forward into the future. 

\section{Direct Dark Matter Searches} 
\par As highlighted above, both spin-independent and spin-dependent direct dark matter searches are now reaching the sensitivity to probe the theoretically well-motivated Higgs-mediated dark matter interactions. The best modern limits on WIMP spin-independent interactions over the entire mass range of $10-1000$ GeV now come from the XENON100 experiment~\citep{2012arXiv1207.5988X}, which has a maximal sensitivity to an elastic scattering cross section of $\sim 10^{-45}$ cm$^2$ at $50$ GeV. As these limits continue to improve, and optimistically close in on a confirmed detection, it will be increasingly important to understand and separate the three components that represent systematic uncertainties that go into determining the dark matter properties. Very broadly, these systematics can be classified as those that arise from the experimental backgrounds in the analysis, those that arise from theoretical uncertainties in the prediction for the WIMP-nucleon cross section, and those that arise from our uncertainty in the distribution of the mass and the velocity distribution of the dark matter. It is probably true that when a WIMP detection is confirmed, the third of these systematics will be the most important systematic that impacts the determination of WIMP properties. For the remainder of this section, I highlight different theoretical and observational aspects of this final systematic, focusing in particular how it affects modern results, and how it can be dealt with more rigorously in the future. 

\subsection{Local dark matter} 
\par The local dark matter density is determined from measurements of the local distribution of stars and their kinematic information. However, extraction of the local dark matter density is rendered difficult because it appears that the dark matter is subdominant to the various baryonic matter components in the Solar neighborhood. Summing up the contributions from low-mass stars, as well as gas from various temperature phases of the interstellar medium, the total mass density of local baryonic matter is approximately $0.1$ M$_\odot$ pc$^{-3}$~\citep{Holmberg:1998xu}. Recent analysis of the kinematics of bright stars finds that the local dark matter density may be up to three times larger than the canonical value of $0.3$ GeV cm$^{-3}$, though these results still systematically depend on the inputs of the analysis and the specific stellar population that is utilized~\citep{Garbari:2011dh,Garbari:2012ff}. Note that even these estimates are still below the local baryonic material by up to a factor of several, and they still carry both significant systematic and statistical uncertainties. Various other analyses that add in constraints from the total Galactic potential find that this uncertainty is reduced, and the mean central value for the dark matter density is slightly larger~\citep{Catena:2009mf,McMillan:2011wd}. However, it is still important to note that these latter estimates are sensitive to the shape and the scale radius of the Milky Way dark matter halo. 

\par The local kinematic measurements above are primarily measuring the ``smooth" distribution of dark matter in the Solar neighborhood. However, from theoretical predictions of dark matter halo formation in cold dark matter cosmology, the distribution of dark matter in the Galactic halo is not smooth, so in principle it is possible that the Sun resides in either a significant local over or under density of dark matter, which may affect the implied constraint on the WIMP elastic scattering cross section. Numerical simulations~\citep{Vogelsberger:2008qb}, as well as analytic models~\citep{Kamionkowski:2008vw}, have begun to address this issue, finding that the probability for the Sun to reside in a significant over density or under density is small, of order $10^{-4}$\%. Further, there are predictions of a dark matter disk in the Galaxy, that may have its origin in the accretion of a massive satellite galaxy that was dragged into the disk by dynamical friction~\citep{Read:2008}. However, analysis of stellar kinematics that extend out beyond a few kilo-parsecs place strong limits on a dark matter disk component in the Galaxy~\citep{Bidin:2010rj}. 

\subsection{WIMP Velocity distribution}
\par The WIMP-nucleon scattering event rate depends in a more phenomenologically interesting manner on the velocity distribution of WIMPs in the halo. Although we have measurements of the distribution of stellar velocities in the disk and in the extended stellar halo of the Milky Way, the only methods that we have available to study the dark matter velocity distribution is through theoretical modeling, numerical simulations, or, most ideally, through direct detection of WIMPs themselves. The mean WIMP event rate scales as $\int d^3 \vec v f(\vec v)/v$, where $f(v)$ is the velocity distribution. This scaling can be simply understood by noting that direct detection experiments are sensitive to the mean WIMP velocity, $\int d^3 \vec v v f(\vec v)$, and at low energies the WIMP-nucleon cross section scales as $1/v^2$. Direct dark matter searches typically assume the so-called standard halo model (SHM) for $f(v)$ to interpret their results in terms of a WIMP mass and cross section. The SHM is an isotropic maxwellian distribution with a cut-off imposed at the local Galactic escape velocity~\citep{Lewin:1995rx}. Translated into position space, the SHM velocity distribution corresponds to an isothermal dark matter density profile. 

\par Although the SHM is useful for calibrating results from different direct detection experiments, it is now becoming clear that the SHM is not the appropriate description of the velocity distribution of dark matter halos in N-body simulations. Indeed, the highest resolution N-body simulations of Milky Way-mass halos find that the velocity distribution differs from the SHM in a several important and interesting ways~\citep{Vogelsberger:2008qb,Kuhlen:2009vh}. The peak of the distribution is broader than is expected from the SHM. Though the physical origin of this is unclear, it may be a reflection of the different dispersions for the different velocity components. Distinct features in the velocity distribution are present due to individual subhalos, and broader, more extended features are apparent out in the power law tail of the distribution that reflect features in energy space~\citep{Vogelsberger:2008qb}. Finally, and probably the most critical for the purposes of direct dark matter detection, the extreme high velocity tail of the distribution appears to be suppressed relative to the SHM. 

\par Although the dark matter velocity distribution is generated from a combination of complicated physical processes that include violent relaxation, phase mixing, and smooth accretion, and the distribution function is certainly neither spherical nor isotropic, it is worthwhile to consider whether the features of the simulated velocity distributions can be understood in the context of simplified theoretical models of the distribution function. In order to gain the best physical intuition, the simplest assumptions to make are that the dark matter velocity distribution is spherical, isotropic, and the system is isolated and in equilibrium. If the dark matter density profile falls off in the outer region as a power law with $r^{-\gamma}$, then by taking the limit of the energy distribution as the binding energy approaches zero it is possible to show that the tail of the velocity distribution will also fall off with a power law index $k$ such the $k = \gamma - 3/2$~\citep{Little:1987}. Note that in this terminology, the commonly used Navarro-Frenk-White profile has $r^{-3}$. This result generally implies that $k$ is determined by the shape of the potential in the outer region where the potential is controlled by the outer slope.

\par Numerical solutions for the velocity distribution show the aforementioned relation between the density profile and the velocity distribution is appropriate for particles within a few percent of the tail of the distribution~\citep{Lisanti:2010qx}. Though the resolution of N-body simulations is not at the level required to fully probe this relation, there are some indications that the power law tail of the velocity distribution is steeper than is predicted by the SHM for the highest resolution halos~\citep{Lisanti:2010qx}. Understanding these properties of the tail of the distribution is particular important for WIMPs in the mass regime of approximately $10$ GeV; a WIMP at this mass scale may be able to reconcile signals reported in a couple of direct detection experiments with sensitivity to low energy nuclear recoils~\citep{Bernabei:2010mq,Aalseth:2010vx}. 

\par The N-body simulations discussed above provide us with highly precise estimates of the velocity distribution in a small number of halos that were re-simulated from halos at larger cosmological volumes. In order to more robustly test the trends that have been seen, it is required to study the velocity distribution of a larger sample of Milky Way-mass halos. Naturally, when extracting simulated halos from a larger cosmological volume (and not re-simulating them at higher resolution, as in the case of the simulated halos discussed above), resolution is an important issue that prohibits a robust determination of the velocity distribution for individual halos. As a concrete example, typical Milky Way-mass halos extracted from large scale simulations~\citep{BoylanKolchin:2009an,Klypin:2010qw} have particle mass about five orders of magnitude larger than those in Refs.~\citep{Vogelsberger:2008qb,Kuhlen:2009vh}. However, what is lost in resolution can be gained by using a larger sample of halos over a wider halo mass range. Using a large sample of dark matter halos from Milky Way-mass to cluster mass scales~\citep{Wu:2012wu}, and stacking the resulting velocity distributions, Mao et al.~\citep{Mao:2012hf} find that the broad properties of the velocity distribution translate more globally in cold dark matter cosmologies. In the Mao et al. study, the stacked velocity distribution is well-described by the following functional form, 
\begin{equation}
f(v) = \exp(-|v|/v_0)(v-v_{esc})^p, 
\label{eq:universal_VDF} 
\end{equation} 
where $v_0$ and $p$ are fitting parameters that depend on the radial position relative to the scale radius. Due to the behavior of the exponential, this distribution implies a wider peak than the SHM, which is a better description of what is found in the simulated distributions. It also better characterizes the power law fall-off of the distribution at high velocities (note that $p \ne k$, where $k$ is defined above as the asymptotic tail of the distribution). Though the theoretical origin of this distribution is not clear, in a manner similar to which the universal origin of dark matter density profiles in cold dark matter simulations is unknown, it may be directly pointing towards global trends in the respective distributions of the different principal components of the velocity distribution. For low mass WIMPs and low threshold detectors, the distribution in Equation~\ref{eq:universal_VDF} implies significant deviations in the WIMP event rate relative to the SHM distribution. 

\par Though in the discussion above I have motivated the effects of the variations in the WIMP velocity distribution on the mean WIMP-nucleon event rate, to which modern experiments are most sensitive, it is certainly true that the velocity distribution also strongly affects different signatures of WIMPs in underground detectors. These include the annual modulation signal, and also further into the future, the signal in directional dark matter detectors. The annual modulation signal is a sensitive function of both the minimal velocity to scatter a nucleus above the detector threshold, and the anisotropy of velocities in the halo~\citep{Fornengo:2003fm}. In addition, directional dark matter detectors may be able to directly determine the anisotropy of the velocity distribution, in addition to the WIMP mass, though present theoretical estimates indicate that the number of events required to cleanly extract the signal is substantial~\citep{Lee:2008jp,Alves:2012ay}.

\subsection{Extracting WIMP properties} 
\par Direct dark matter searches are clearly in the midst of the ``discovery" phase, attempting to extract the WIMP signal from background sources. The ultimate goal of direct dark matter searches is of course not only to detect the dark matter, but also to extract information on its properties such as the mass and the cross section.  In addition to examination of the particle properties, it will be interesting to determine if we can use the detection of WIMPs to understand properties of the Galactic halo. Examples of these properties are the local density, velocity distribution, as well as any more detailed features in these distributions that may result from substructures or streams. Understanding the extent to which it is possible to extract both particle and astrophysical properties will certainly require comprehensive modeling of both of these components.  

\par Some recent theoretical work has focused on understanding whether different direct detection experiments are consistent with a dark matter signal, given that they have different values for their thresholds and different backgrounds to deal with~\citep{Fox:2010bz,Gondolo:2012rs}. Formalisms like this will need to be further expanded upon when looking forward towards the potential ``detection" phase of direct detection. Ultimately, a rigorous statistical formalism is required to model the data using input from both the astrophysical and the particle physics components. Initial studies that focused on extracting WIMP properties in direct detection experiments did so using phenomenological models for the velocity distribution~\citep{Green:2008rd}. Due to the larger WIMP event rate and the lack of form factor suppression in the cross section, these studies brought to light the important result that lower mass WIMPs ($< 100$ GeV) are much better constrained in direct detection experiments than larger mass WIMPs ($>100$ GeV). 

\par Strigari and Trotta~\cite{Strigari:2009zb} extended the aforementioned analyses by developing a bayesian method for extracting WIMP properties, including uncertainties in a wide range of Galactic model parameters. These parameters include the local dark matter density, the circular velocity, the dark matter halo mass, as well as the properties of the Milky Way disk. For a one ton-year exposure with liquid xenon, they find that the mass of a fiducial $50$ GeV WIMP can be determined to a precision of less than approximately 25\%. For WIMPs greater than $100$ GeV, the mean event rate distribution will only be able to provide a lower bound on the WIMP mass. Combining different detector targets holds promise for further improving the determination the WIMP mass and cross section~\citep{Pato:2010zk}, in particular using experiments with very different nuclear masses. Even including uncertainties in Galactic model parameters, determination of the WIMP mass and cross section is unbiased when properly marginalizing over the uncertainties in the Galactic model parameters. This result also holds when using parameterizations of the velocity distribution that are closely related to the SHM, though this result on the bias of the parameter reconstruction clearly needs to be explored further for different models of the velocity distribution. 

\subsection{Astrophysical backgrounds}
\par Finally, on the topic of direct dark matter searches, it is interesting to point out that ultimately direct dark matter searches will not be zero background experiments. Indeed, this reflects the fact that the descendants of modern direct dark matter detection experiments were developed for the purposes of solar neutrino detection~\citep{Cabrera:1984rr}. Due to the vector interactions with neutrons in the nucleus, for neutrinos with energies of less than approximately $10$ MeV, the neutrino-nucleus cross section is enhanced by the factor of the square of the mass number. This enhancement is similar to the coherent enhancement of spin-independent WIMP-nucleon interactions. At the lowest detectable recoil energies of $\sim 3$ keV in liquid xenon, solar neutrinos will provide a background for experiments at approximately the 1 ton scale~\citep{Monroe:2007xp}. Larger mass detectors at the scale of approximately $20-100$ ton will be sensitive to atmospheric and diffuse supernova neutrinos over all recoil energy ranges of interest~\citep{Strigari:2009bq}. For a $100$ GeV WIMP, this corresponds to a spin independent cross section of approximately $10^{-48}$ cm$^2$. Though it will make extraction of the WIMP signal more difficult at this scale, it likely will not be an ``irreducible" background, because the energy spectrum of WIMPs is distinct from each of the neutrino signals. For a large enough sample of events, it will be in principle possible to do spectral analyses of each of the different sources. And of course even farther into the future, it would clearly be possible to distinguish the WIMP signal from the isotropically-distributed atmospheric and supernova signals using dark matter detectors with directional sensitivity. 

\section{Going forward} 
\par We are now right in the beginning of the experimental era in particle dark matter searches. Though there have been hints of detections, at different levels of plausibility, it is important to bear in mind that we are now just reaching the sensitivity, through both direct and indirect detection methods, to probe the most well-motivated models from cosmology and particle physics. At this stage, the Fermi-LAT dSph analysis has been the first experiment to achieve robust sensitivity to thermal relic WIMP dark matter, doing so in the mass regime $10-25$ GeV. Direct dark matter searches are now just approaching the theoretically motivated Higgs-mediated regime, and are now probing well-motivated regimes of the supersymmetric parameter space. 

\par Over the course of the next several years, the sensitivities of direct and indirect searches will continue to improve, which will certainly improve the modern limits, and hopefully even reveal interesting signals. From the point of view of gamma-ray studies, at least seven years of Fermi-LAT data is expected, clearly statistically improving the two year results from the Fermi-LAT discussed above. However, there are also reasons to believe that the sensitivity will improve more rapidly than is expected just from photon counts alone. This is particularly true from the perspective of the dSph searches. First, through galaxy surveys that are now coming online, such as Pan-STARRS~\footnote{http://pan-starrs.ifa.hawaii.edu/public/}, the Dark Energy Survey (DES)~\footnote{http://www.darkenergysurvey.org/}, and even further into the future the Large Synoptic Survey Telescopes (LSST)~\footnote{http://www.lsst.org/lsst/}, we are certain to discover more faint satellite galaxies around the Milky Way, in a manner similar to the methods used by the Sloan Digital Sky Survey (SDSS) to discover nearly a dozen new ultra-faint satellites. In particular in the southern sky, only a handful of satellites are now known, and these are only the brightest objects that have been known for nearly a century since their discovery on photographic plates. As new objects are detected, kinematic analysis of their constituent stars will provide their dark matter masses, and adding these objects to the Fermi-LAT analysis will certainly improve the limits on particle dark matter. It is still even possible that new surveys will find a massive, nearby satellite, similar to the SDSS discovery of the satellite Segue 1. In addition to the potential for detection of new satellites that can be added to the analysis, with an extended mission lifetime for Fermi-LAT, and new cleaned analysis of the gamma-ray data and a better understanding of the background, it will be possible to obtain more information from the gamma-ray data. This data will be of particular interest over the next several years, when extended source models for the satellite galaxies will be used in the Fermi-LAT analysis pipeline. With this combination of improvements, it will certainly be possible to reach the thermal relic cross section scale for dark mater masses of $100$ GeV over the next several years. Finally, the proposed Cherenkov Telescope Array (CTA)~\footnote{http://www.cta-observatory.org/}, which is expected to begin operation near 2017, will extend thermal relic dark matter limits to much higher masses, beyond the TeV scale, with better angular resolution than Fermi-LAT. Combining results from the Fermi-LAT and CTA, within the next decade we will cover fully the thermal relic dark matter parameter space above the TeV mass scale.

\begin{theacknowledgments}
I would like to especially thank Barbara Szczerbinska for organizing the CETUP* workshop, Bhaskar Dutta for organizing the dark matter program, as well 
as all the participants of the dark matter program for stimulating discussions. I thank Yao-Yuan Mao for discussions on topics that I cover in these proceedings. 
\end{theacknowledgments}

\end{document}